# Near field transmission using Hermite-Gaussian modes

Chenxi Zhu, *Senior member, IEEE*
Wireless Research Lab, Lenovo Research, Beijing, China
zhucx1@lenovo.com

*Abstract*—RF transmission in line-of-sight near field based on Hermite-Gaussian (HG) modes is developed. Multiple HG modes are transmitted and received using rectangular antenna arrays to form the basic modes and dimensions for MIMO transmission. Beam steering can be achieved by manipulating the antenna arrays with 3D rotation in the desired EM field. The beam parameters are optimized to minimize the size of the antennas. Simulation is performed for a 300GHz system with free space channel model. Spectrum efficiency up to 294.3bps/Hz can be achieved with 36 HG modes and cross-polarization.

## I. INTRODUCTION

MIMO communication has extended into near field as the antenna size continues to grow in the 6G era [1]. Near field refers to the region where the electromagnetic radiation cannot be treated as (superposition of) planar waves, and intrinsic structure of the EM wave needs to be considered. While near field in 6G may be up to 100 meters or more, near field is also prominent in short range line-of-sight (LOS) communications. In the optical field, structured light refers to light wave that maintains certain spatial structure during propagation [2]. Broadly speaking, this includes any EM wave (including RF) with spatial inhomogeneity of amplitude, phase and other parameters. In free space the wave equation needs to satisfy the Helmholtz equation, the solution of which gives rise certain modes of EM propagation. Structured light can be used for short range transmission in the near field before the structure dissipates into planar waves in the far field. Examples of structured light include Laguerre-Gaussian (LG) beam, Hermite-Gaussian (HG) beam, Ince-Gaussian (IG) beams, and Kummer beam [3]. These structures have long been used in the optical field to analyze laser beam inside the cavity and in free space and optical fiber, and for precision measurement and high-resolution imaging. Among various structured lights, LG modes (OAM) have been used broadly for communication, first in optical, and more recently in RF domain [4]-[8]. Different OAM modes can carry separate information for parallel transmission. Another member of the Gaussian beam family is Hermite-Gaussian modes. Like LG modes, HG modes also form a set of orthonormal basis for paraxial wave and can be used to analyze and synthesize propagation in the near field [9]. Conversion between the two bases is possible [10]. Because HG modes have square or rectangular cross sections, it is easy to represent a beam with a rectangular profile. HG modes have been used, though less frequently than LG modes, in free space optical communication [11][12]. In the millimeter wave domain, low order HG modes have been generated using phase plate [13] or patch antenna [14]. In [15] an all-electric time-domain mode-division-multiplexing (TD-MDM) method based on HG modes was proposed as a form of CDMA for wireless transmission.

In this paper we explore HG modes for spatial multiplexing in RF transmission and use multiple HG modes to carry parallel data streams. We construct a set of parallel channels based on HG modes and use them as degree-of-freedom or dimensions for MIMO transmission. We also develop a scheme to steer the beam in any direction relative to the antenna array.

## II. HERMITE-GAUSSIAN (HG) MODES

Hermite-Gaussian modes are solutions of the paraxial Helmholtz equation in the Cartesian coordinate [3]. Assume the direction of transmission is Z. Solving the paraxial Helmholtz equation gives a set of orthonormal solutions, each characterized by a pair of integers $(l,m)$:

$$HG_{l,m}(x,y,z) = \sqrt{\frac{1}{2^{l+m-1}\pi l! m!}} \frac{1}{w(z)} H_l\left(\frac{\sqrt{2}x}{w(z)}\right) H_m\left(\frac{\sqrt{2}y}{w(z)}\right) \exp\left(-\frac{x^2+y^2}{w^2(z)}\right) \cdot \exp\left(-i\frac{k(x^2+y^2)}{2R(z)}\right) \exp\left(i\psi_{l,m}(z)\right) \exp(-ikz) \quad (1)$$

$H_l(\cdot)$ is the Hermite polynomial (the physicist type) of order $l$. $H_l(x)$ and $H_m(y)$ determines the shape of the beam in the X and Y directions respectively. The curvature of wavefront is $\frac{1}{R(z)} = \frac{z}{z^2 + D_{Ray}^2}$. The Rayleigh distance $D_{Ray} = \frac{\pi w_0^2 n}{\lambda}$ is determined by the radius of the beam waist $w_0$ on the focal plane, and $w(z) = w_0\sqrt{1 + \frac{z^2}{D_{Ray}^2}}$ is the beam radius at distance $z$ from the focal plane. We take the refraction index $n = 1$ for air, and $\lambda$ is wavelength, $k = \frac{2\pi}{\lambda}$. The term $\psi_{l,m}(z) = (1 + l + m)\arctan\left(\frac{z}{z_{Ray}}\right)$ is Gouy's phase. Each $HG_{l,m}(x,y,z)$ is normalized and represents a distinct mode traveling in the Z direction. Different modes maintain orthogonality while traveling through space. The electric field can be decomposed into a summation of different HG modes

$$E(x,y,z) = \sum_{(l,m)} E_{l,m} HG_{l,m}(x,y,z) \quad (2)$$

where $E_{l,m}$ represents the strength of mode $(l,m)$ and is related to the transmission power $P_{l,m} = E_{l,m}^2$. The $n$-th order Hermite polynomial, given by

$$H_n(x) = (-1)^n e^{x^2} \frac{d^n}{dx^n} e^{-x^2} \quad (3)$$

has leading coefficient $2^n$ and $n$ zeros on the real axis. High order Hermite polynomials can be derived from lower order polynomials through the recursion $H_{n+1}(x) = 2xH_n(x) - 2nH_{n-1}(x)$. Hermite polynomials of different orders are mutually orthogonal with the weight function $e^{-x^2}$:

$$\int_{-\infty}^{\infty} H_m(x)H_n(x)e^{-x^2}dx = \sqrt{\pi}\,2^n n!\,\delta_{m,n} \quad (4)$$

where $\delta_{m,n}$ is the Kronecker delta function. Each $HG_{l,m}$ is a unique propagation mode representing a degree of freedom in space, and different HG modes are mutually orthogonal and can carry different data streams in parallel for spatial multiplexing. In phase $\exp\left(-i\frac{k(x^2+y^2)}{2R(z)}\right)\exp\left(i\psi_{l,m}(z)\right)\exp(-ikz)$, the first term is the phase variation on the beam cross section, and the second and third terms only depend on z and represent the phase in the wavefront. In a cross section, the phase is constant in a circle $x^2 + y^2 = r^2$. Unlike LG modes, there is no angular dependent term, so a HG mode does not carry any orbital angular momentum. Compared with a LG mode which has a spiral wavefront, a HG mode has a pseudo-spherical wavefront. The phases of different HG modes only differ slightly through Gouy's phase. The wavefronts of different HG modes have almost the same shape and do not diverge like LG modes. Spin angular momentum manifests as polarization and does not affect the beam profile or the shape of the wavefront. When cross-polarized antenna is used, the two perpendicular polarization directions can be manipulated independently as in traditional cross-polarized MIMO systems. In the following discussion we will not distinguish between unidirectional or cross-polarization most of the time.

### III. TRANSMISSION OF HG MODES

Because the electric field of HG modes is perpendicular to the direction of propagation, HG modes can be generated using a planar array consisting of uniform electric antennas. In Sec III and IV, TX and RX antennas are collimated and point to each other. Excitation of the electric field by the antennas in the XY-plane according to the wave equation (2) generates HG modes propagating in the Z direction. The many (unidirectional or dual-polarized) transmitting antennas are arranged in uniform square grid as in Figure 1. For the TX (or RX) array, we assume there are $(2N_t + 1)^2$ (or $(2N_r + 1)^2$) antennas (from $-N_t$ to $N_t$ or $-N_r$ to $N_r$ in the X and Y directions). The distance between adjacent antennas is $d$. Varying $d$ changes the sampling rate in the spatial domain and affects the sidelobe but does not affect the shape of the main lobe, while $d$ should not be too large to cause significant overlap between the two. Depending on the modes used, rectangular antenna array can also be used.

Suppose the direction of transmission is Z, and the origin is set at the focal point of the beam. When the TX antenna is placed perpendicular to Z at $z = z_T$, the electric field for mode $(l, m)$ on the antenna plane is $E_{l,m}HG_{l,m}(x, y, z_T)$. To transmit normalized modulation symbol $s_{l,m}$ using mode $(l, m)$, antenna $(i, j)$ at location $(x_i, y_j, z_T)$ sends the signal $t_{l,m}(i, j) = E_{l,m}HG_{l,m}(x_i, y_j, z_T)s_{l,m}$. For cross-polarized antenna, a pair of data symbols $(s_{l,m}^+, s_{l,m}^-)$ can be transmitted.

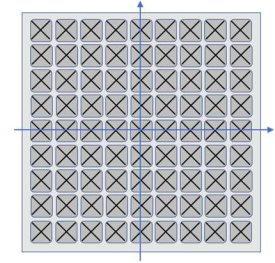

Figure 1. Square (rectangular) uniform planar array for transmission and reception.

The coefficients for all the transmission antennas $W_{l,m} = \left[HG_{l,m}(x_i, y_j, z_T), -N_t \leq i, j \leq N_t\right]$ form the TX precoder or the spatial transmission filter for mode $(l, m)$. When multiple modes $M$ are transmitted, antenna $(i, j)$ transmits their summation:

$$t(i,j) = \sum_{(l,m) \in M} E_{l,m}HG_{l,m}(x_i, y_j, z_T)s_{l,m} \quad (5)$$

Figure 2 shows the power profiles for some individual modes on the TX array plane. The parameters are given in Table 1. For $l \geq l', m \geq m'$, the beam profile of $(l, m)$ is larger than $(l', m')$. If $(l_{max}, m_{max})$ is the largest mode number that can be transmitted by a rectangular antenna array, the total number of modes that can be transmitted is $(l_{max} + 1)(m_{max} + 1)$.

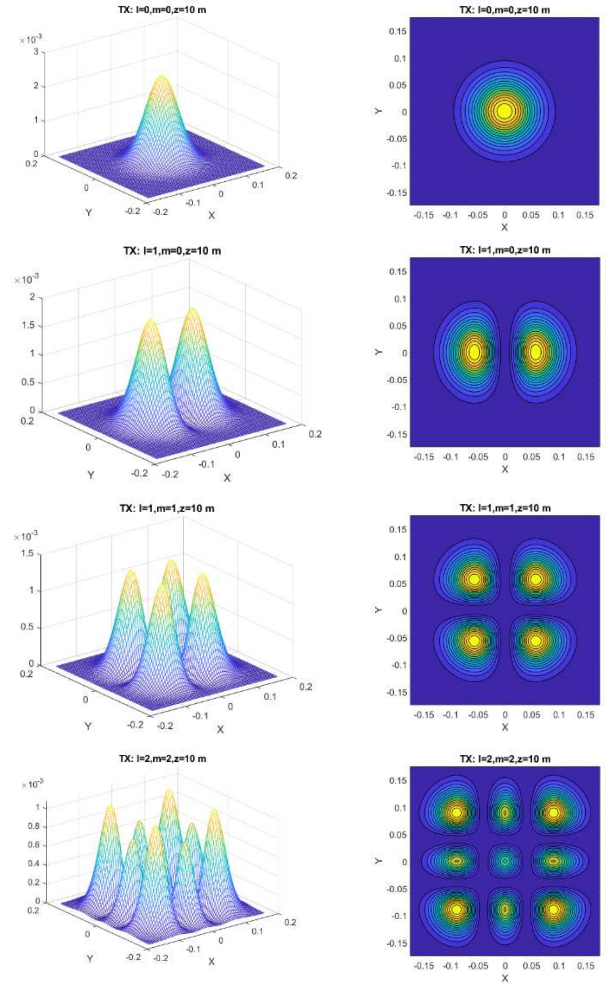

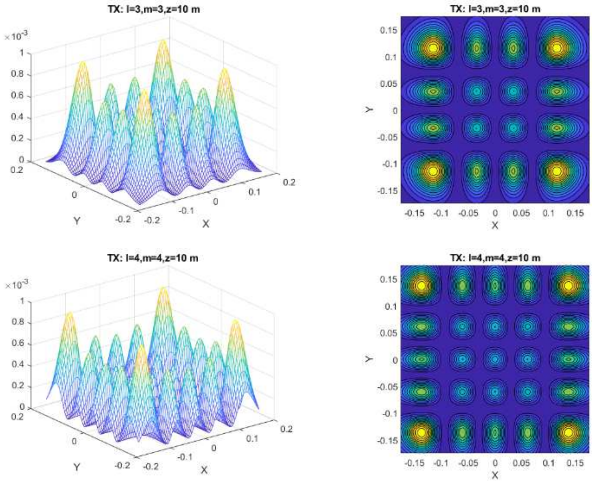

Figure 2. TX Power profiles on the TX antenna plane. From top to bottom: mode (0,0), (1,0), (1,1), (2,2), (3,3), (4,4), (5,5).

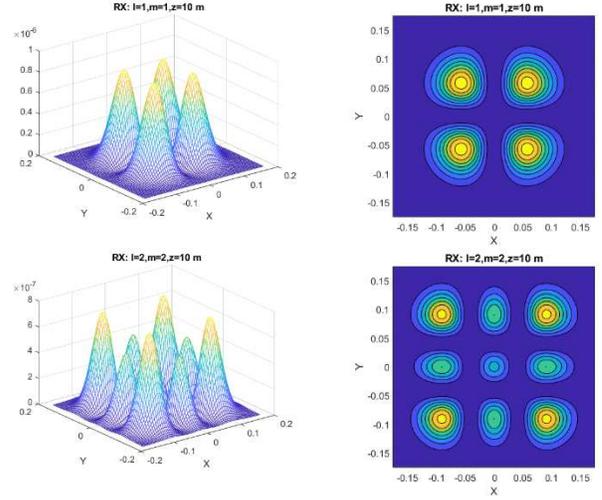

Figure 3. RX power profiles on the RX antenna plane. From top to bottom: mode (1,1), (2,2).

## IV. RECEPTION OF HG MODES

The uniform antenna array in Figure 1 can be used for receiving HG modes. As the transmission from all the TX antennas form HG modes propagating through space, the signals received by the RX antennas capture a cross section of the traveling beam on the RX plane. Figure 3 shows the received signal profiles on the RX antenna for some HG modes. The received signal resembles the profile of the transmitted signal very well (with reduced power), indicating the qualities of the HG modes generated and received are both very good. Because different modes maintain orthogonality while traveling along the Z axis, $HG_{l,m}^*(x, y, z_R)$ can be used as the receiving filter by the RX antenna at $z_R$. To recover the symbol sent in mode $(l, m)$, the receiver applies the coefficient $HG_{l,m}^*(x_i, y_j, z_R)$ to signal $r(x_i, y_j, z_R)$ received on the antenna $(i, j)$ summing over all the RX antennas:

$$\begin{aligned}
u_{l,m} &= \sum_{-N_r \leq i,j \leq N_r} HG_{l,m}^*(x_i, y_j, z_R) r(x_i, y_j, z_R) \\
&= \sum_{-N_r \leq i,j \leq N_r} \sum_{(l',m') \in M} HG_{l,m}^*(x_i, y_j, z_R) E_{l',m'}^r \\
&\quad \cdot HG_{l',m'}(x_i, y_j, z_R) s_{l',m'} \\
&\cong \sum_{(l',m') \in M} \int_{-N_r d}^{N_r d} \int_{-N_r d}^{N_r d} E_{l',m'}^r HG_{l,m}^*(x_i, y_j, z_R) \\
&\quad \cdot HG_{l',m'}(x_i, y_j, z_R) s_{l',m'} dxdy \\
&\cong \sum_{(l',m') \in M} \int_{-\infty}^{\infty} \int_{-\infty}^{\infty} E_{l',m'}^r HG_{l,m}^*(x, y, z_R) \cdot \\
&\quad HG_{l',m'}(x, y, z_R) s_{l',m'} dxdy = E_{l,m}^r s_{l,m}
\end{aligned} \quad (6)$$

where $E_{l',m'}^r$ is the received signal amplitude for mode $(l', m')$ on the RX plane. The two approximations (limited antenna size and discrete antenna elements) lead to nonorthogonality between different modes, and their effects will be evaluated numerically in the following sections. To recover the transmitted symbols, the received symbols for all the modes $\{u_{l,m}, (l,m) \in M\}$ can be further processed using standard signal processing schemes such as the MMSE algorithm.

An effective channel can be constructed based on the transmitted HG modes. Let $G^p$ be the channel matrix between the TX and the RX antennas. Mode-based effective channel matrix $H^{mod} = \left[h_{(l_r,m_r),(l_t,m_t)}^{mod}\right]_{|M| \times |M|}$ can be computed as

$$h_{(l_r,m_r),(l_t,m_t)}^{mod} = HG_{l_r,m_r}^* G^p HG_{l_t,m_t} \quad (7)$$

with the mode-specific transmission and receiving filters applied to the TX and RX ends. The coefficient $h_{(l_r,m_r),(l_t,m_t)}^{mod}$ is the effective channel for mode $(l_t, m_t)$ when the TX and RX modes are the same, and interference from mode $(l_t, m_t)$ to mode $(l_r, m_r)$ otherwise. Due to limited antenna size and discrete antenna elements, different modes are not strictly orthogonal. This leads to the off-diagonal elements in $H^{mod}$. Each mode is an independent degree of freedom (DOF). The DOF for the point-to-point transmission is $|M|$ for unidirectional and $2|M|$ for cross-polarization. With the effective channel $H^{mod}$, two transmission schemes can be developed using the standard MIMO algorithms. The first is to decompose $H^{mod}$ using singular value decomposition (SVD) to generate a set of orthonormal precoding vectors on top of the HG modes for transmission. This can be called SVD-based transmission scheme. It should be noted that the SVD is not performed on the physical channel $G^p$ and the resulting precoding vectors are not the native SVD precoders between the TX and RX antennas. The receiver can use the corresponding left singular vectors and the MMSE algorithm for reception. The second scheme is to transmit different data streams using the HG modes directly as in (6), ignoring their cross interference on the TX side. No additional TX precoder is applied on top of the HG modes. This can be called the HG-mode based transmission scheme. This is useful when the off-diagonal terms are relatively small, and the interference can be dealt with by the receiver using the MMSE algorithm. Although suboptimal compared with the SVD-based scheme, this approach does not require measurement of cross-mode interference. For simplicity, the transmission power is evenly split among all the used modes in both transmission schemes.

## V. ANTENNA DESIGN AND PARAMETER OPTIMIZATION

The TX and RX antennas need to match the cross section of the beam to be efficient. We optimize the beam parameters to reduce the antenna size. The beam cross section is proportional to the beam radius $w(z)$, which increases with the distance $z$ from the focal plane. When the TX and RX antennas are $D_{TR}$ apart, changing the position of the focal plane can change the size of the antennas. If the focal plane is placed at one end (TX or RX), the beam (thus the antenna) at the other end will be much larger. When the TX and RX antennas have the same constraint, we can place them symmetrically ($-z_T = z_R = 0.5 D_{TR}$) with respect to the focal plane so they have the same size. We can optimize the beam waist $w_0$ on the focal plane to reduce the beam radius $w(z_T)$ on the TX/RX plane. The beam radius at the TX (or RX)

$$w_{RX}(z_R) = w_0\sqrt{1 + \frac{z_R^2}{D_{Ray}^2}} = w_0\sqrt{1 + \frac{z_R^2 \lambda^2}{\pi^2 w_0^4}} \quad (8)$$

can be minimized by

$$w_0^{opt} = \sqrt{\frac{\lambda z_R}{\pi}}, \quad (9)$$

leading to the optimal beam radius on the TX/RX plane

$$w_{TX}^{opt}(z_T) = w_{RX}^{opt}(z_R) = \sqrt{\frac{\lambda D_{TR}}{\pi}} = \sqrt{2} w_0^{opt} \quad (10)$$

The Rayleigh distance is $D_{Ray}^{opt} = z_R$. This puts the TX and RX right at the Rayleigh distance from the focal plane. By minimizing $w_{TX}^{opt}$ (or $w_{RX}^{opt}$), we reduce the antenna size for given HG modes or increase the number of modes that can be fitted onto a given antenna. The physical size of the antennas depends on the $w_{RX}^{opt}$ as well as the modes used. We can evaluate the antenna size by the energy captured. Figure 4 shows the efficiency of square antennas of different sizes for modes $l = m$. The antenna size ($2s \times 2s$) is normalized with respect to the beam radius $w$ on the antenna plane. For $s = w$, only modes (0,0) and (1,1) can be received with ≥50% power. For $s = 1.5w$, modes up to (3,3) can be received with ≥45% power. For $s = 2.2w$, 36 HG modes ($0 \leq l', m' \leq 5$) can be received with ≥70% power.

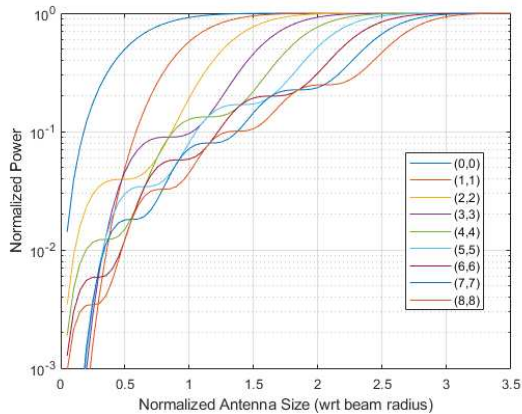

Figure 4. Normalized power of different antenna size.

The power profile for individual mode is very uneven (Figure 2). The picture changes when multiple modes are transmitted together, and the power profile becomes more even in a rectangular area.

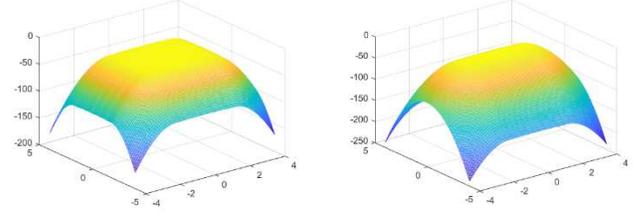

Figure 5. TX power profile (in logarithm) of 36 modes ($0 \leq l, m \leq 5$, left) and 6 modes ($0 \leq l \leq 5, m = 0$, right).

Figure 5 shows the combined power profile for 36 modes ($0 \leq l, m \leq 5$) and 6 modes ($0 \leq l \leq 5, m = 0$). The dimension is normalized with respect to $w$. All modes are sent with equal power. Most of the power is emitted from a square ($|x|, |y| \leq 2.2w$, 36 modes) or a rectangular ($|x| \leq 2.2w$, $|y| \leq 0.25w$, 6 modes) area. The power is relatively uniform within this area, and drops rapidly out of this area. This power pattern can be met by placing PAs uniformly in this area. Different antenna size and different mode numbers are possible in the X and Y dimensions. As a comparison, all OAM modes have circular profiles and require circular antennas, making them less flexible than HG modes.

## VI. SIMULATION RESULTS

We simulate a point-to-point system operating at 300GHz. The simulation assumptions are listed in Table 1.

Table 1. Simulation assumptions.

| Carrier frequency $f_c$ | 300 GHz |
| --- | --- |
| Channel bandwidth | 2000 MHz |
| Transmission power | -6 dBm |
| Transmission distance | 20 m |
| Beam waist ($w_0^{opt}$) on the focal plane | 0.056 m |
| Beam radius ($w_{TX/RX}^{opt}$) on TX/RX plane | 0.080 m |
| TX/RX antenna array size | 0.35m x 0.35m |
| TX/RX antenna configuration | 71x71, interspacing 5 mm |
| Channel model | Frii's free space model |
| Antenna model | 3GPP TS38.901 [18] |
| RX noise figure | 8 dB |
| HG modes used | From (0,0) to (5,5) |
| Polarization | Unidirectional- or cross-polarized |
| Power allocation | Equal power for all modes |
| Modulation and coding scheme | 5G NR modulation and coding up to QAM64 |

Figure 6 shows the SNRs for 36 modes of unidirectional transmission sorted in descending order for the SVD-based and HG-mode based schemes. The TX and RX antennas are perfectly aligned in the direction of transmission. The SNRs of the two schemes are almost identical. This shows good orthogonality between different HG modes, and cross mode interference is insignificant. Applying SVD to $H^{mod}$ did not bring any additional gain. The 5G NR modulation and coding scheme is used to compute the spectrum efficiency and the total throughput. For unidirectional transmission, both schemes achieve spectrum efficiency of 171.7 bps/Hz and of 343.4 Gbps. When cross-polarization is used, the total number of modes doubles to 72. Despite 3dB power reduction for each mode/polarization, the increased degree of freedom brings the spectrum efficiency to 294.3 bps/Hz and throughput to 588.6 Gbps for both schemes.

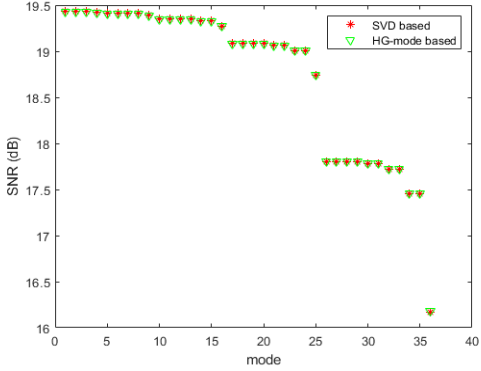

Figure 6. SNRs for SVD-based and HG-mode based schemes for 36 modes, unidirectional polarization, perfect alignment.

## VII. BEAM STEERING

So far we have assumed the TX antenna and the RX antenna are perfectly aligned with each other. This limits the application to a fixed point-to-point link. The transmission will become more flexible if the alignment can be relaxed. In [16], the direction of OAM beam can be adjusted for misaligned UCA antenna, but the steering angle is limited. In [17], IR beam is steered optically using a pair of reflection gratings for indoor communication. We develop an all-electric approach for beam steering without limit to the steer angle. Take TX for example. Because the Hermite-Gaussian mode equation is developed under the paraxial approximation, it is not easy to divert the beam by applying phase shifts to the antennas without significant error. Instead of taking a traditional signal-based approach on a 2D array, we take a 3D EM field-based approach. Since (1) gives the EM field of HG modes in 3D, we can regenerate the EM field from each of the antennas by using their 3D coordinates. We can tilt the antenna in the 3D EM field of a beam traveling in the Z direction to get their transmission coefficients. The beam still travels in the Z direction, but from the viewpoint of the TX antenna it is tilted from the broadside. Define the function $(x', y', z') = R_{(\theta_x, \theta_y)}(x, y, z)$ to rotate a point along the X-axis by $\theta_x$ and along the Y-axis by $\theta_y$. Suppose we want to transmit in the direction $(\theta_x, \theta_y)$ from the broadside. If we tilt the original antenna plane $P_{tx}$ (in the XY plane) by $(-\theta_x, -\theta_y)$, antenna $i$ at $(x_i, y_i, z_i)$ will be rotated to new coordinate $(x_i', y_i', z_i') = R_{(-\theta_x, -\theta_y)}(x_i, y_i, z_i)$ on a new plane $P_{tx}'$. For mode $(l, m)$ each antenna $i$ gets a new transmission coefficient $HG_{l,m}(x_i', y_i', z_i')$ corresponding to the electric field of the mode propagating in the Z direction. We now have a mode $(l, m)$ propagating away from the antenna array $P_{tx}'$ by $(\theta_x, \theta_y)$. If we tilt the antenna plane back to $P_{tx}$ but keep the coefficient $HG_{l,m}(x_i', y_i', z_i')$, the beam will be steered in the direction $(\theta_x, \theta_y)$ from the Z axis. Figure 7 shows the TX power profile for mode (2,2) on the TX antenna tilted (30º, 30º), and the received power profile on the RX antenna facing the direction of transmission. The profile on TX antenna (top) is stretched, but the profile on the RX antenna (bottom) resembles the RX profile sent from perfectly aligned TX antenna (Figure 3 bottom). This proves the signal generated from the tilted TX antenna is still in the original direction. Because no approximation is made in this process, there is no limit to the steering angle. Similarly, when the RX antenna plane is not facing the direction of transmission, the weight of the receiving filter can be computed with a rotation. However, the stretched beam profile may fall off the limited antenna area (Figure 7, top) and cause distortion to the signal pattern. High order modes will be impacted more due to their large profiles. This not only degrades the signal but leads to loss of orthogonality between different HG modes.

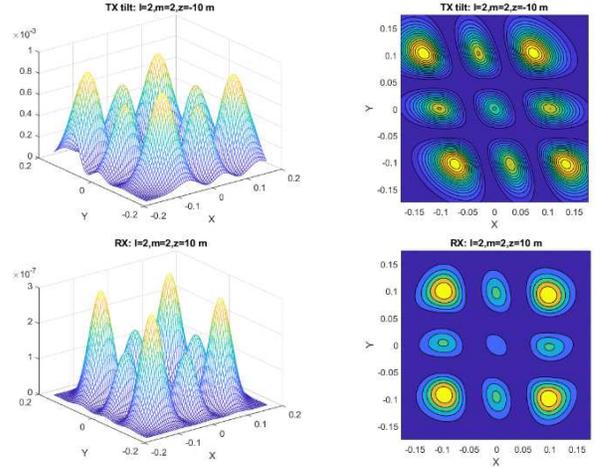

Figure 7. Beam profile on TX (tilted (30º, 30º), top) and RX (untilted, bottom) antenna for mode (2,2).

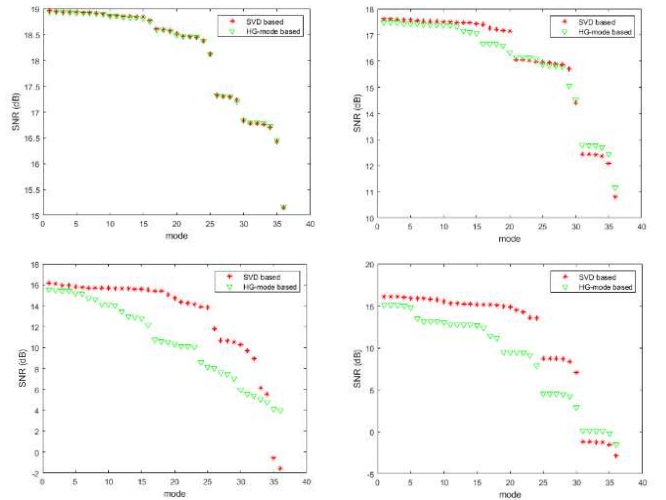

Figure 8. SNRs when TX beam is tilted by (15º,0º, top left), (30º,0º, top right), (30º,30º, bottom left), (45º,0º, bottom right).

We study the effect of beam steering with simulation. The set up is the same as in Table 1. The TX beam is steered in different directions from the broadside of the TX antenna, while the RX antenna always faces the direction of transmission. Figure 8 shows the SNRs for the SVD-based and the HG-mode based transmission schemes as the transmitter sends its beam in different angles. The spectrum efficiencies are shown in Figure 9. SNR and spectrum efficiency both drop as the beam is steered more from the broadside. However, the degradation is more significant for the HG-mode based scheme. As the beam is more skewed, the beam profile is distorted more and cut off more from the TX antenna. This causes more nonorthogonality between different HG modes, especially for large mode numbers. While the SVD-based scheme can eliminate the cross-mode interference from the TX side through orthogonalization, the HG-mode based scheme cannot. This leads to severe degradation as the beam is tilted more, and the gap between the two schemes widens. To improve the performance, we can use only those HG modes with good SNRs or use the water-filling algorithm to allocate transmission power among them. Since these are standard techniques in MIMO communication, we will not pursue them further here.

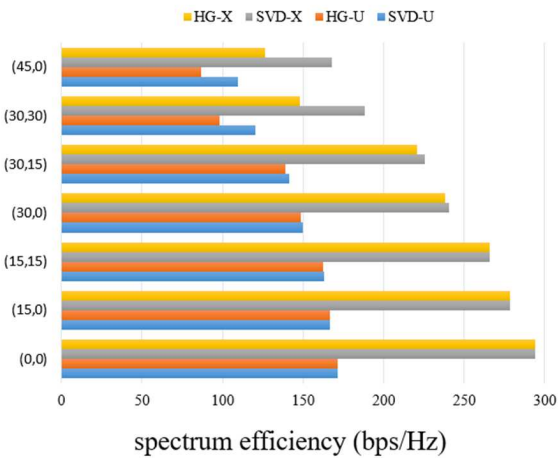

Figure 9. Spectrum efficiency for TX beam tilted in different directions ($\theta_x, \theta_y$). HG: HG-mode based. SVD: SVD-based. U: unidirectional. X: cross-polarized.

## VI. Conclusion and future work

We have developed a near field transmission scheme based on Hermite-Gaussian modes. By using rectangular antenna arrays to generate and receive HG modes, spatial multiplexing using multiple HG modes to transmit parallel data streams can be realized. We also optimized the beam parameters to minimize the antenna size.

In this paper we used the ideal free space channel model. As a next step we will study the effect of scattering with a more realistic channel model. With our method of beam steering, although the TX or RX antennas do not need to be perfectly aligned in the direction of transmission, the direction needs to be known to both sides. In a real system misalignment may occur, and mechanical or electrical correction is needed. Deep learning has been used in OAM-based system from air turbulence compensation [19] to mode detection [20]. We plan to explore deep learning for misalignment detection and correction for HG modes.

Steerable beam opens the possibility for a transmitter to transmit simultaneously to more than one receivers in different directions, achieving so-called multi-user MIMO (MU-MIMO) transmission. When the receivers are sufficiently apart, the transmitter can superimpose the signals meant for all of them and transmit together on the TX antenna, sending different beams in different directions toward their targets. We will explore MU-MIMO in detail in subsequent works.